\documentclass{nature}
\usepackage[utf8]{inputenc}
\usepackage[T1]{fontenc}
\usepackage{graphicx}
\usepackage{xcolor}
\usepackage{physics}
\usepackage{bm}
\usepackage{amsfonts}
\makeatletter
\let\saved@includegraphics\includegraphics
\AtBeginDocument{\let\includegraphics\saved@includegraphics}
\renewenvironment*{figure}{\@float{figure}}{\end@float}
\makeatother

\title{Beam dynamics induced by the quantum metric of exceptional rings}

\author{Zhaoyang Zhang$^{1\ast\dagger}$, Ismaël Septembre$^{2,3\ast\dagger}$, Zhenzhi Liu$^{1\dagger}$, Pavel Kokhanchik$^2$, Shun Liang$^1$, Fu Liu$^1$, Changbiao Li$^1$, Hongxing Wang$^1$, Maochang Liu$^4$, Yanpeng Zhang$^1$, Min Xiao$^{5}$,
Guillaume Malpuech$^{2\ast}$, Dmitry Solnyshkov$^{2,6\ast}$}

\begin{document}
\maketitle

\begin{affiliations}
\item Key Laboratory for Physical Electronics and Devices of the Ministry of Education \& Shaanxi Key Lab of Information Photonic Technique, School of Electronic Science and Engineering, Faculty of Electronics and Information, Xi'an Jiaotong University, Xi'an 710049, China
\item Institut Pascal, PHOTON-N2, Universit\'e Clermont Auvergne, CNRS, Clermont INP,  F-63000 Clermont-Ferrand, France
\item Naturwissenschaftlich-Technische Fakultät, Universität Siegen, Walter-Flex-Straße 3, 57068 Siegen, Germany
\item International Research Center for Renewable Energy and State Key Laboratory of Multiphase Flow in Power Engineering Xi’an Jiaotong University, Xi’an 710049, China
\item Department of Physics, University of Arkansas, Fayetteville, Arkansas, 72701, USA
\item Institut Universitaire de France (IUF), 75231 Paris, France
\end{affiliations}

* marks corresponding authors, E-mails: zhyzhang@xjtu.edu.cn, ismael.septembre@uni-siegen.de, guillaume.malpuech@uca.fr, dmitry.solnyshkov@uca.fr\\
$\dagger$: these authors contributed equally to this work.

\section*{Abstract}
\begin{abstract}
Topological physics has broadened its scope from the study of topological insulating phases to include nodal phases containing band structure singularities. The geometry of the corresponding quantum states is described by the quantum metric which provides a theoretical framework for explaining phenomena that conventional approaches fail to address. 
The field has become even broader by encompassing non-Hermitian singularities: 
in addition to Dirac, Weyl nodes, or nodal lines, it is now common to encounter exceptional points, 
exceptional or Weyl rings, and even Weyl spheres. They give access to fascinating effects that cannot be reached within the Hermitian picture. However, the quantum geometry of non-Hermitian singularities is not a straightforward extension of the Hermitian one, remaining far less understood. Here, we study experimentally and theoretically the dynamics of wave packets at exceptional rings stemming from Dirac points in a photonic honeycomb lattice. First, we demonstrate a transition between conical diffraction and non-Hermitian broadening in real space. Next, we predict and demonstrate a new non-Hermitian effect in the reciprocal space, induced by the non-orthogonality of the eigenstates. We call it transverse non-Hermitian drift, and its description requires biorthogonal quantum metric. The non-Hermitian drift can be used for applications in beam steering. 
\end{abstract}

\maketitle

Non-Hermitian Hamiltonians, despite appearing contradictory with the very postulates of standard quantum mechanics, are very useful and efficient~\cite{Moiseyev2011} to describe the distribution of losses among states while preserving coherence. This is particularly relevant for photonic systems, where losses are inherently present. Generally, the eigenvalues of the non-Hermitian Hamiltonians are complex and the corresponding eigenstates are non-orthogonal. Exceptional points are characterized by the degeneracy of the complex eigenvalues associated with a topological singularity and by the coalescence of the eigenvectors.
These points have been extensively studied in the past few years, especially in photonics~\cite{Miri2019}. Close to an exceptional point, the eigenvalues scale as an integer power root of any perturbation, which notably leads to high sensitivity~\cite{Wiersig2014,Wiersig2016,hodaei2017enhanced,chen2017exceptional,park2020symmetry} (potentially infinite, but limited by the noise~\cite{duggan2022limitations}). It can be traced back to the quantum metric describing the quantum distance between eigenstates \cite{Wiersig2022}: the quantum metric diverges close to exceptional points~\cite{Liao2021}. This is similar to the divergence of the quantum Fisher information~\cite{Zanardi2007} at phase transitions.

\begin{figure}
\centering
\includegraphics[width=0.9\linewidth]{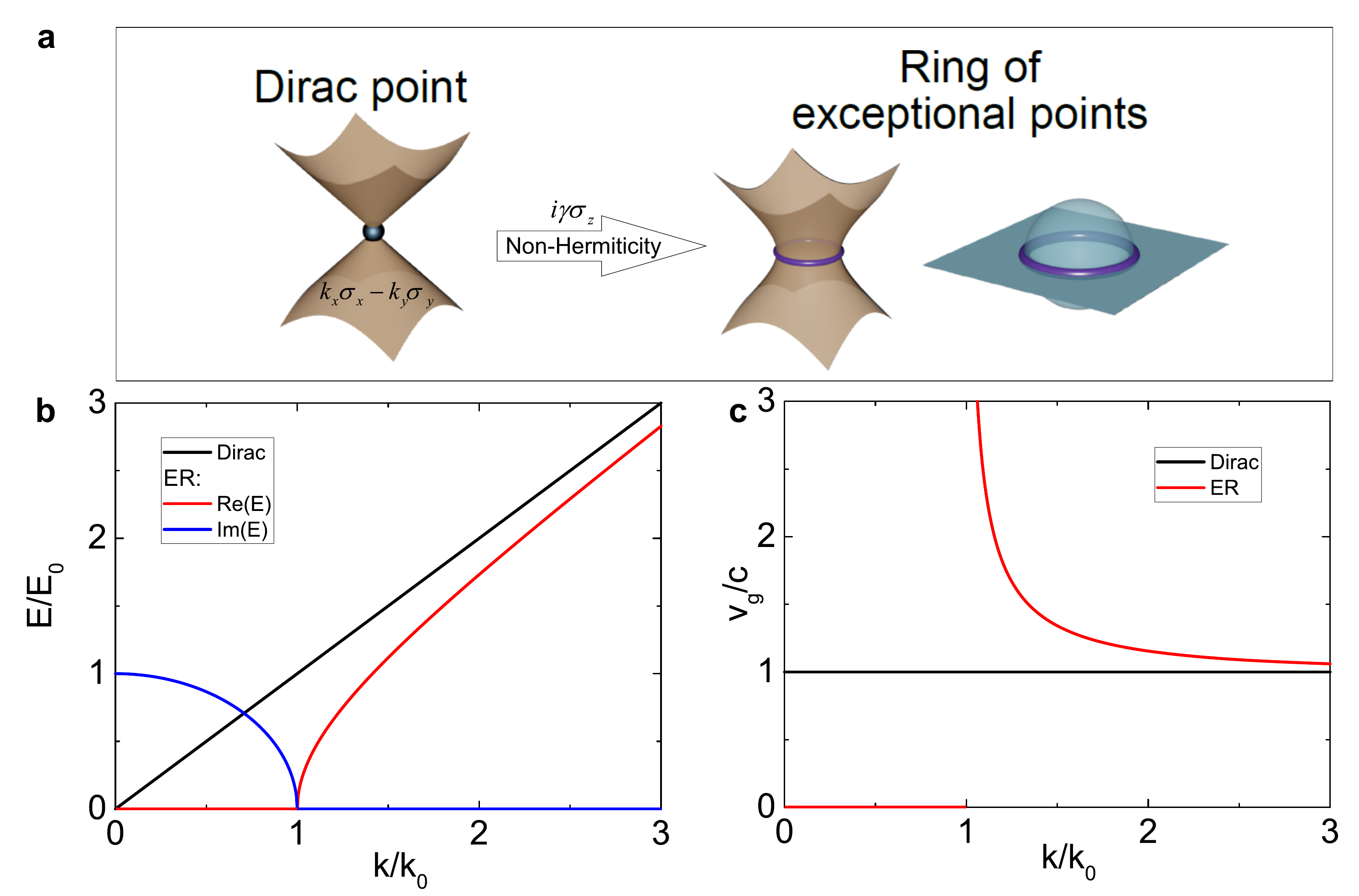}
\caption{\textbf{Exceptional ring and its properties} 
\textbf{a} (Sketch) Dispersion relation of the Dirac point transformed into an exceptional ring by non-Hermiticity  (real part in brown, imaginary part in blue).
\textbf{b} Dispersion as a function of wave vector $k$ (black - Hermitian case, red - $\Re(E)$, blue - $\Im(E)$). \textbf{c} Group velocity as a function of $k$ (black/red - Hermitian/non-Hermitian cases).
 \label{fig1}}
\end{figure}

In two-band systems, exceptional points always appear in pairs connected by real Fermi arcs, where the real part of the energy is degenerate. Interestingly, these pairs of points can form a ring (see the sketch in Fig.~\ref{fig1}a), called a ring of exceptional points~\cite{zhen2015spawning} or simply an exceptional ring~\cite{xu2017weyl,cerjan2019experimental}. Each pair of points on the ring's diameter is linked by a Fermi arc: the real part of the eigenvalues is degenerate within the whole ring. 
After the first observations in photonic systems in two~\cite{zhen2015spawning} and three~\cite{cerjan2019experimental} dimensions, rings of exceptional points have been in the focus of increasing interest \cite{xu2017weyl,yoshida2019exceptional,xu2022observation,liu2022experimental,li2023exceptional} far beyond the initial field of non-Hermitian lattices  \cite{Leclerc2024}. While the properties of the dispersion and the group velocity were analyzed quite extensively~\cite{cerjan2019experimental,liu2022experimental}, the dynamic behavior of wavepackets at the exceptional rings has not been deeply studied yet. This dynamics is important, because it provides a better microscopic understanding of the macroscopic transport phenomena (as was the case for the anomalous Hall effect \cite{Sundaram1999}), and because of the possible practical applications.

The dynamics of wave packet in lattices is often determined by the Berry curvature~\cite{Sundaram1999,Bliokh2007}, or, more generally, by the quantum geometric tensor that also includes the quantum metric~\cite{Torma2018,Bleu2018effective,leblanc2021universal,Torma2023essay,hu2024generalized}. This led to an important question of the choice of the quantum metric~\cite{provost1980riemannian} for the description of non-Hermitian systems~\cite{Brody2013,brody2013biorthogonal}. Indeed, the quantum metric and more generally the quantum distance as a measure of a scalar product (overlap) between the states implies the use of both bra and ket vectors. In Hermitian systems, the bra vector is just a conjugate of the ket vector, but in non-Hermitian systems, due to the non-orthogonality of the eigenstates, the bra eigenvectors should be found as dual to the ket eigenvectors (like the basis of the reciprocal lattice forms a dual set with respect to the direct lattice). In our case, these dual vectors are the left eigenvectors of the  Hamiltonian~\cite{moiseyev1978resonance,Moiseyev2011,brody2013biorthogonal}. Depending on the problem and on the type of experiment, one can be tempted to use either only right eigenvectors, or both left and right~\cite{solnyshkov2021quantum,Hu2023}.
Despite the hot debate, where different theoretical approaches were proposed~\cite{Brody2013,brody2013biorthogonal}, and several experiments suggested to test them~\cite{Ye2024}, the question is still open~\cite{hu2024generalized}. In experiments, the quantum metric measured so far has been the standard one~\cite{gianfrate2020measurement,Yu2019} (based on the right eigenvectors only), even in non-Hermitian systems~\cite{Liao2021,Cuerda2024}.

In this work, we use a well-established honeycomb photonic lattice in rubidium atomic vapors in the regime of electromagnetically-induced transparency (EIT)~\cite{PhysRevLett.74.666} in order to unveil the role of the quantum metric. We introduce non-Hermiticity ~\cite{Feng2023} by inducing different losses on A and B lattice sites, which results in the transformation of Dirac points into exceptional rings.  
This transition is evidenced by the qualitative change in the probe dynamics propagating through the vapor cell: we observe a transition from conical diffraction to non-Hermitian spatial broadening. The scaling of the size of the wave packet with the non-Hermiticity magnitude is quantitatively determined by the quantum metric. Next, we demonstrate both experimentally and theoretically the presence of a non-Hermitian counterpart of the anomalous Hall effect. We observe a drift of the probe wavepacket in the reciprocal space. This drift occurs not only in the direction of the gradient of the imaginary part of the energy \cite{hu2024generalized}, towards the center of the exceptional ring, but more surprisingly, also in the transverse direction. We call this effect the transverse non-Hermitian drift and demonstrate that it is entirely due to the non-orthogonality of the eigenstates and requires the biorthogonal quantum metric for its description. The resulting non-Hermitian reciprocal-space dynamics can be applicable to beam steering~\cite{zhao2019non,heck2017highly}.

We begin by a general theoretical description of a Dirac point and an exceptional ring and their properties, important for the understanding of the  dynamics of the wave packet. We take an effective Hamiltonian of the form:
\begin{equation}\label{HamNH}
    H = k_x \sigma_x - k_y \sigma_y + i\gamma \sigma_z = \left ( \begin{array}{cc}
        i\gamma & ke^{i\theta} \\
        k e^{-i\theta} & -i\gamma 
    \end{array} \right ),
\end{equation}
which in our case describes a particular Dirac point of a honeycomb lattice with loss staggering~\cite{Feng2023} (neglecting the average losses). Here $k=\sqrt{k_x^2+k_y^2}$ is the wave-vector norm, $\theta=\arctan{k_y/k_x}$ is the wave-vector angle ($\bm{k}$ is measured from the Dirac point), and a single parameter $\gamma$, which represents the loss imbalance (obtained by nondimensionalization  as 
$\gamma=\gamma_0/\hbar c$, where $c$ is the celerity of wave packets at the Dirac point and $\gamma_0$ is the non-Hermiticity of the dimensional Hamiltonian). This Hamiltonian is non-Hermitian $H \neq H^\dagger$ because of the loss staggering term $i\sigma_z$. Its eigenvalues  $E_\pm = \pm \sqrt{k^2-\gamma^2}$ are represented as a function of wave vector $k$ in Fig.~\ref{fig1}b (Hermitian limit: black line, exceptional ring: $\Re (E)$ -- red line, $\Im (E)$ -- blue line). The radius of the exceptional ring is $k_0=\gamma$ and we normalize the energies by $E_0=E_{Herm}(k_0)=\gamma$. The square root scaling of both real and imaginary parts close to $k=k_0$ is clearly visible. This dispersion relation gives rise to a modified group velocity: while it is constant for a Dirac point, it diverges as $v_g=\hbar^{-1}\Re \partial E/\partial k\sim (k-k_0)^{-1/2}$ close to the exceptional ring, as shown in Fig.~\ref{fig1}c (Hermitian: black line, exceptional ring: red line). The group velocity is zero inside the exceptional ring, which strongly affects the real-space behavior of the wave packets. The eigenstates of the Hamiltonian in Eq.~\eqref{HamNH} are also important for the understanding of the behavior of the system. Their explicit expressions are provided in Methods and their behavior is shown in Extended Data Fig.~\ref{EDFpseudoNOmetric}.

\begin{figure}
\centering
\includegraphics[width=0.9\linewidth]{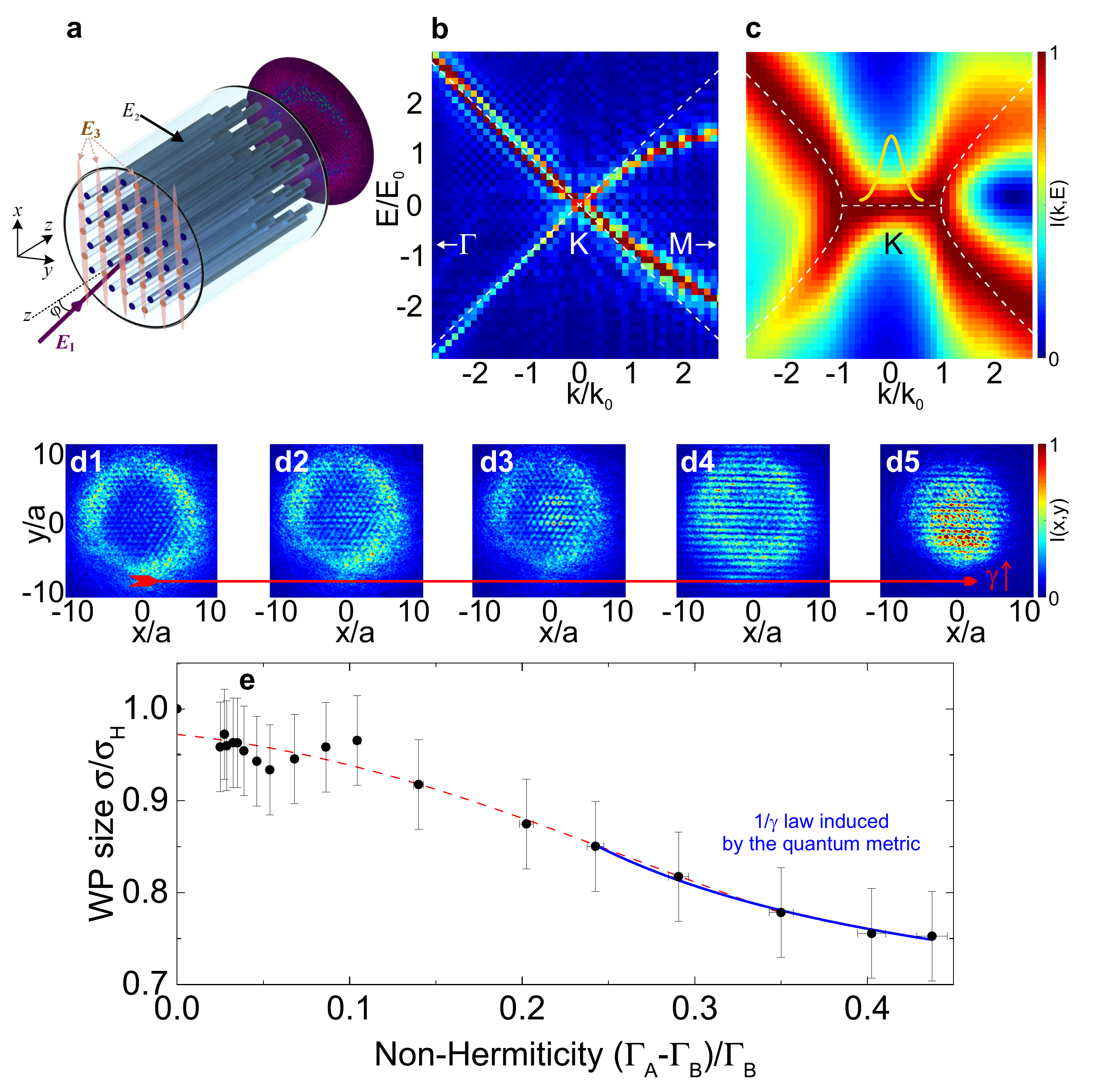}
\caption{\textbf{Configuration of the experiment and real-space results.} 
\textbf{a} The scheme of the experiment with an atomic vapor cell: $\boldsymbol{E_1}$ is the probe beam, $\boldsymbol{E_2}$ field creates a honeycomb lattice, $\boldsymbol{E_3}$ field provides a staggered imaginary potential. \textbf{b,c} Dispersion of the photonic lattice in the Hermitian (b - Dirac) and non-Hermitian (c - Exceptional ring) cases, zoomed on the K point. A yellow Gaussian represents the initial wave packet for panels d1-d5. \textbf{d1-d5} Real-space images of the output beam exhibiting conical diffraction and its suppression in a non-Hermitian transition ($a$ is the lattice parameter).  \textbf{e} Width of the wave packet in real space $\sigma$ as a function of the non-Hermiticity $(\Gamma_A-\Gamma_B)/\Gamma_B$ (black dots -- experiment, red and blue lines -- theory), with $\sigma_H$ -- the width in the Hermitian case. \label{fig2}}
\end{figure}

To study the dynamics of the wave packets at exceptional rings experimentally, we create a photonic honeycomb lattice with staggered losses and send a probe at one of the Dirac points of this lattice. The experimental setup is depicted in Fig.~\ref{fig2}a (see Methods for more details). It consists of a rubidium vapor cell. The honeycomb photonic lattice is optically induced by a hexagonal coupling field $\boldsymbol{E_2}$ (formed by 3-beam interference) under the condition of EIT. The EIT effect produced for a Gaussian probe beam $\boldsymbol{E_1}$ by the lattice beams $\boldsymbol{E_2}$ makes the susceptibility inversely related to the coupling-field intensity under certain frequencies of the two involved fields \cite{ PhysRevLett.129.233901, zhang2020observation, PhysRevLett.132.263801}. The frequency detunings of the probe and coupling fields are set as $\Delta_1 = -50$ MHz and $\Delta_2 = -80$ MHz, respectively (see Extended Data Fig.~\ref{figS1}). They are defined as the difference between the frequency gap of the two energy levels driven by $\boldsymbol{E_i}$ and its frequency (Extended Data Fig.~\ref{figS1}). The staggered losses are created by an extra field $\boldsymbol{E_3}$ (2-beam interference), which covers and modulates the A sites \cite{PhysRevLett.132.263801}, allowing to control the losses on these sites relatively to the B sites. The probe beam is captured by a camera behind the output plane of the vapor cell. The detailed simulations of the susceptibility experienced by the probe $\boldsymbol{E_1}$ can be found in Ref.~\cite{Feng2023} and in Methods (Extended Data Fig.~\ref{figS2}). 

The dispersion of the resulting photonic lattice close to the Dirac point in the Hermitian and non-Hermitian cases obtained numerically is shown in Fig.~\ref{fig2}b,c, respectively. The linear Dirac cone is clearly visible in Fig.~\ref{fig2}b, and the flat real part of the exceptional ring together with the square root behavior at its edges can be seen in panel Fig.~\ref{fig2}c. The linear dispersion of the Dirac cone leads to the well-known phenomenon of conical diffraction~\cite{Hamilton1837,Lloyd1837,berry2006conical}: a Gaussian beam is converted into a ring, because all components propagate with the same group velocity (Fig.~\ref{fig1}c, black line), but in different directions. This is observed in our experiments when the loss staggering is off (Fig.~\ref{fig2}d1). We then  turn on $\boldsymbol{E_3}$ field and change its frequency to smoothly increase the staggered losses, which allows to control the non-Hermiticity of the Hamiltonian as $\gamma\sim (\Gamma_A-\Gamma_B)/\Gamma_B$ (with $\Gamma_{A,B}$ the losses on $A$ and $B$ sites, respectively -- see Methods). The associated non-Hermitian transition of the output beam between a hollow and a filled circle  is shown in panels d1-d5 presenting selected images showing the probe intensity at the output of the vapor cell for increasing values of non-Hermiticity. This transition is actually a topological one, because the non-zero wave function phase winding associated with conical diffraction~\cite{berry2006conical,Zhang2019} switches to zero. The conical diffraction disappears because the components inside the exceptional ring exhibit zero group velocity (Fig.~\ref{fig1}c, red line).
To analyze the behavior of the wave packet quantitatively, we study its root mean square width $\sigma=\sqrt{\langle r^2 \rangle-\langle r\rangle^2 }$. Figure~\ref{fig2}e shows experimentally measured wave packet size (normalized to the value observed in the Hermitian case $\sigma_H$) as a function of non-Hermitian staggering $(\Gamma_A-\Gamma_B)/\Gamma_B$, together with two theoretical curves showing the analytical solutions valid in the limit of small (red) and large (blue) non-Hermiticity. It is in the latter limit that the biorthogonal quantum metric of non-Hermitian systems plays an important role, as we show below.

The solution of the time-dependent Schrödinger equation for a given initial condition $\ket{\psi(0)}$ can generally be written as
$\ket{\psi(t)}=\sum_i c_i \exp(-iE_i t/\hbar)\ket{\psi_i}$, where the eigenvalues $E_i$ of the stationary Schrödinger equation $H\ket{\psi_i}=E_i\ket{\psi_i}$ can be, in general, complex. The coefficients $c_i$ form the decomposition of the initial wave function over the basis formed by the right eigenvectors $\ket{\psi_i}$, but if this basis is not orthogonal (because of the non-Hermiticity), their calculation by $c_i=\braket{\phi_i}{\psi(0)}$ requires the use of the dual basis formed by the left eigenvectors $\bra{\phi_i}H=\bra{\phi_i}E_i$. The overlaps $c_i$ are determined by the quantum metric:  $|c_i|^2=\cos^2 s=\cos^2 \int\limits_{\ket{\psi(0)}}^{\bra{\phi_i}}  \sqrt{g^i_{\mu\nu}dk_\mu dk_\nu}$, where $s$ is the quantum distance between the specific pair of states involved in each case. Since they involve not only right, but also left eigenvectors (which are the ones forming the dual basis), the metric tensor $g^i_{\mu\nu}$ that needs to be used in this expression is necessarily the biorthogonal one~\cite{Brody2013,brody2013biorthogonal}, containing both left $\bra{\phi_i}$ and right $\ket{\psi_i}$ eigenvectors
\begin{equation}\label{QMbiorth}
    g_{\mu\nu}^i =\Re\left[ \left \langle \frac{\partial\phi_i}{\partial \mu} \middle | \frac{\partial\psi_i}{\partial \nu} \right \rangle - \left \langle \frac{\partial\phi_i}{\partial \mu} \middle | \psi_i \right \rangle \left \langle \phi_i \middle | \frac{\partial\psi_i}{\partial \nu} \right \rangle\right],
\end{equation}
with the normalization condition $\braket{\phi_i}{\psi_i}=1$. The explicit expressions of its elements for the Hamiltonian~\eqref{HamNH} are provided in Methods.

The probability density associated with the wave function $\ket{\psi(t)}$ can be explicitly written for our two-band system ($i=\pm$) as 
\begin{multline}
     p(t)=|\ket{\psi(t)}|^2=|c_+|^2\braket{\psi_+}{\psi_+}e^{-2\Im E_+t/\hbar}+c_+^* c_-\braket{\psi_+}{\psi_-}e^{-i\Re (E_--E_+)t/\hbar}\\
    +c_-^* c_+ \braket{\psi_-}{\psi_+}e^{-i\Re (E_+-E_-)t/\hbar}+|c_-|^2\braket{\psi_-}{\psi_-}e^{-2\Im E_-t/\hbar}
    \label{sol}
\end{multline}
The difficulty of the problem that has led to the debates on the choice of the metric tensor is partially linked with the fact that such complete expressions involve scalar products of two different types.  The first type is the scalar product  of left and right eigenvectors in the decomposition coefficients $c_i$. The second type is the scalar product of right eigenvectors only. Terms such as $\braket{\psi_-}{\psi_+}$ appear due to the non-orthogonality of right eigenstates (Extended Data Fig.~\ref{EDFpseudoNOmetric}b, magenta curve). The calculation of the above expression in terms of quantum metric elements requires therefore \textit{both} the biorthogonal quantum metric and the standard one (both are provided in Methods). Here, we focus mostly on the contribution and the consequences of the biorthogonal quantum metric, in order to demonstrate its physical importance.

\begin{figure}
\centering
\includegraphics[width=0.99\linewidth]{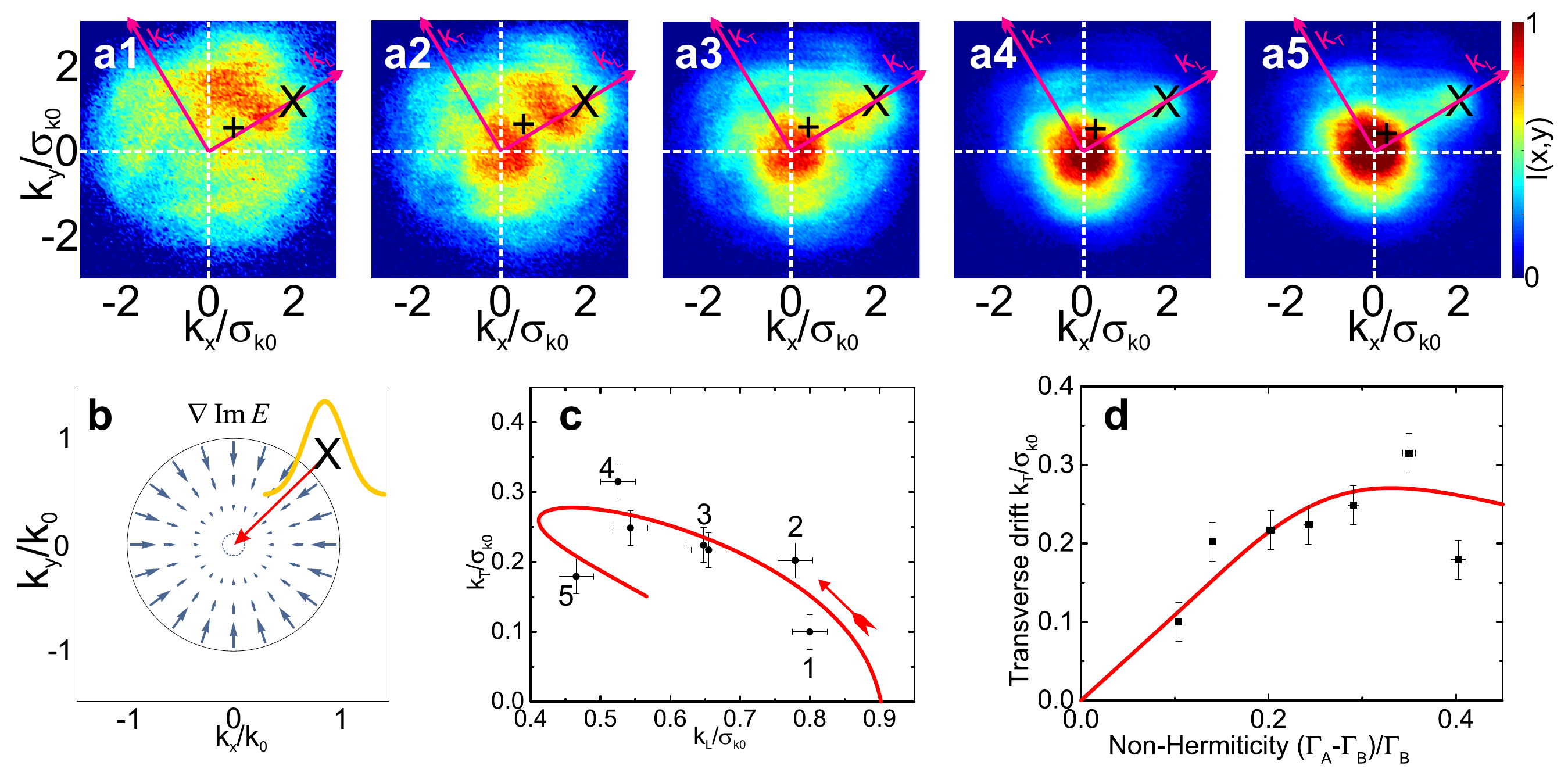}
\caption{\textbf{Longitudinal and transverse non-Hermitian drift in reciprocal space.} 
\textbf{a1}-\textbf{a5} Images of the output wave packet in the reciprocal space with increasing non-Hermiticity (experiment), "X" marks the initial position of the wave packet and "+" marks the center of mass of the snapshot, longitudinal $k_L$ and transverse $k_T$ axes are shown in magenta; \textbf{b} 
Sketch of the gradient of the imaginary part of the energy (blue arrows), responsible for the longitudinal drift (red arrow); \textbf{c}
Position of the center of mass in the reciprocal space at fixed $t$ for different non-Hermiticity (numbers correspond to panels a1-a5, arrow marks the non-Hermiticity increase); \textbf{d} Transverse (anomalous) non-Hermitian drift as a function of non-Hermiticity. In \textbf{c,d}: black dots -- experiment, red curve -- theory, error bars correspond to experimental uncertainty. \label{fig3}}
\end{figure}

In the case of a wave packet centered at the Dirac point (yellow Gaussian curve in Fig.~\ref{fig2}c) in presence of a large non-Hermiticity ($k\ll \gamma$), all energies are purely imaginary (Fig.~\ref{fig1}b) and no interference effects can arise ($\Re (E)=0$). The wave packet size in the reciprocal space can be calculated as $ \sigma_k (t) = \sqrt{ \int\limits_{0}^\infty  k^2 p_k (k,t) 2\pi k dk}$, where $p_k(k,t)=|\psi(k,t)|^2$ is the probability distribution in the reciprocal space. The dependence on $k$ includes an initial Gaussian wave packet $G(k)$ of a size $\sigma_{k0}$ and the k-dependent overlaps. The non-orthogonal terms of the right basis appearing in Eq.~\eqref{sol} are non-zero $\braket{\psi_\pm}{\psi_\mp}=k/\gamma$, but cancel out in this case, because of the conjugation. Therefore, the probability density reduces to $p(k,t)=N(t)G(k)(|c_+|^2e^{2\Im E t}+|c_-|^2e^{-2\Im E t})$, where $N(t)$ is a normalization coefficient. For an initial spinor $\ket{\psi(0)}=(1,0)^T$ required for conical diffraction, using the biorthogonal quantum metric which is approximately constant $g_{kk}\approx 1/4\gamma^2$ for $k\approx 0$, the integral defining the quantum distance reduces to a product $s\approx\sqrt{g_{kk}}k$, and we obtain one of the overlaps with the dual basis as $|c_-|^2\approx g_{kk} k^2$. The other is obtained similarly: 
$|c_+|^2\approx 1-g_{kk} k^2$. This allows us to calculate analytically the width of the wave packet in the reciprocal space for $\gamma t\ll 1$ as
\begin{equation}
    \sigma_k(t)\approx\sigma_{k0}\left(1-8g_{kk}\sigma_{k0}^2\gamma t\right)
\end{equation}
Since there is no phase evolution because $\Re E=0$, the \textit{real-space} wave packet width that we actually measure grows as $\sigma(t)\approx\sigma_0(1+8g_{kk}\sigma_{k0}^2\gamma t)$. In our experiment, the propagation time $t$ is fixed by the vapor cell length, so we study $\sigma$ as a function of $\gamma$: $(\sigma-\sigma_0)/\sigma_0\sim g_{kk}\gamma\sim \gamma^{-1}$ (Fig.~\ref{fig2}e, blue line).  For smaller values of non-Hermiticity, the analytical solution (Fig.~\ref{fig2}e, red dashed line) is discussed in Methods. The predicted metric-induced power law $g_{kk}\gamma\sim\gamma^{-1}$ is in a very good agreement ($R^2=0.989$) with the experimental data of Fig.~\ref{fig2}e.

We now study the dynamics of wave-packet center of mass in the reciprocal space with its initial center of mass position $\bm{k}_0$  away from the Dirac point and even outside the exceptional ring $k_0>\gamma$. Selected experimental images of the final wave packet distribution for increasing non-Hermiticity are shown in Fig.~\ref{fig3}a. The behavior of the center of mass (marked by "+") can be decomposed into two effects: longitudinal and transverse drift. The positions of the center of mass in the reciprocal space obtained from experimental snapshots are shown in rotated longitudinal-transverse axes (with $\bm{k}_L\parallel \bm{k}_0$, $\bm{k}_T\perp  \bm{k}_0$, shown in Fig.~\ref{fig3}a in purple) in Fig.~\ref{fig3}c with black dots. The longitudinal drift (along $k_L$) is explained by the gradient of the imaginary part of the eigenvalues $\nabla \Im (E)$ sketched in Fig.~\ref{fig3}b with blue arrows (the initial position is shown with a yellow Gaussian curve): the components with longer lifetime dominate at longer times, which brings the wave packet to the Dirac point position $k_{L}\to 0$. This is a finite-size effect pointed out recently~\cite{solnyshkov2021quantum,hu2024generalized}.

The transverse drift corresponds to the motion along $k_T$, perpendicular to the imaginary part gradient. It is visible in the five panels a1 to a5, and its dependence on non-Hermiticity is shown separately in Fig.~\ref{fig3}d). Interestingly, it appears only because of the interference of the non-orthogonal terms in Eq.~\eqref{sol} involving the products $c_-^* c_+ \braket{\psi_-}{\psi_+}$ with complex time-dependent exponents, which add up constructively or destructively depending on the sign of $k_T$. We are interested in the origin of this dependence on the transverse wave vector $k_T$, which can be traced back to the azimuthal quantum metric as follows. As an example, one of the overlaps can be written as $|c_+|=|\braket{\phi_+}{\psi}|\approx s$, and for small $k_T$,
$|c_+|\approx \int\limits_\infty^{k} \sqrt{g_{kk}dkdk}+\int\limits_0^{\theta} \sqrt{g_{\theta\theta}d\theta d\theta}\approx s_l+\sqrt{g_{\theta\theta}}k_T/k$. The contribution linear in $k_T$  is therefore $|c_+|_T=\sqrt{g_{\theta\theta}}k_T/k=k_T/2\sqrt{k^2-\gamma^2}$. 
Other terms are also present in the full expression, but many of them cancel out, giving the product $c_-^* c_+ \braket{\psi_-}{\psi_+}_T=-k_T\gamma g_{\theta\theta}/k^2=-k_T\gamma/4(k^2-\gamma^2)$.
This allows to find an explicit analytical expression for the probability density:
\begin{equation}
    p_k(\bm{k},t)=N(t)G(k)\left(1+2\sin^2\left(\frac{E t}{\hbar}\right)\frac{(k_T+\gamma)\gamma}{k^2-\gamma^2}\right)
    \label{proden}
\end{equation}
This expression is valid for all $\bm{k}$. Outside of the ring, the second term in the parenthesis describes  the interference between the non-orthogonal terms. The term linear in $k_T$ increases the probability density at positive $k_T$ thus leading to a displacement of the k-distribution of the wave packet (transverse wave packet drift). Several time snapshots of the probability density for a fixed non-Hermiticity are plotted in Extended Data Fig.~\ref{EDF1snap}. The distribution given by Eq.~\eqref{proden} perfectly coincides with a full numerical solution of time-dependent Schrödinger equation (up to numerical precision). We also plot the probability density $p_k(\bm{k},t)$ at a fixed time for different values of non-Hermiticity in Extended Data Fig.~\ref{EDF2gam}. The transverse drift appears in these figures as a vertical shift of the wave packet.

Using the probability density~\eqref{proden}, we obtain the position of the center of mass as $\langle \bm{k}(t)\rangle=\iint \bm{k}p_k(k,t)d^2k$. It is plotted in Fig.~\ref{fig3}c as a red curve, and its $k_T$ projection is plotted separately in panel d. In both cases, the evolution time $t$ is fixed by the cell length, and we vary the non-Hermiticity $\gamma$. The trajectories of the center of mass are also plotted in Extended Data Figs.~\ref{EDF1snap},\ref{EDF2gam} as a white line.  The experimental trends are very well described by the theory, and in particular the anomalous non-Hermitian transverse drift, fully determined by the azimuthal biorthogonal quantum metric $g_{\theta\theta}$ (see Extended Data Fig.~\ref{EDFpseudoNOmetric}d).

We note that the "standard" Hermitian quantum metric, based on the right-right eigenvectors, has a different expression ($g_{\theta\theta}^{Herm}=1/4$). It does not allow to correctly calculate the overlaps $c_\pm$, does not provide a positively-defined probability density in Eq.~\eqref{proden}, and thus does not allow to reproduce the exact solution of the Schrödinger equation and the experimental data. However, the Hermitian metric is still required to calculate the overlaps $\braket{\psi_-}{\psi_+}$. We conclude that both tools ($g_{ij}$ and $g_{ij}^{Herm}$) should be used properly to get the correct result.

We have demonstrated experimentally the transition from Hermitian conical diffraction at a Dirac point to non-Hermitian wavepacket broadening, and shown that this broadening is controlled by the radial component of the quantum metric. We have also demonstrated a drift of the wave packet in the reciprocal space with a particularly interesting transverse component, controlled by the azimuthal  biorthogonal quantum metric. This experiment demonstrates that the correct description of non-Hermitian systems should employ left and right eigenvectors and the biorthogonal quantum metric tensor. Our work could have potential applications in beam steering for integrated photonic circuits and beyond~\cite{ouyang2007light,heck2017highly,Luk2021,meng2023topological}.

\begin{methods}

\noindent{\textbf{Experimental configuration and data treatment.}}
The weak Gaussian probe beam (wavelength $\approx$ 795.0 nm) is from an external-cavity diode laser (ECDL1) and horizontally polarized. Three vertically polarized coupling beams (wavelength $\approx$ 795.0 nm) from ECDL2 intersect in a small angle of $0.4^{\circ}$ with each other to form the hexagonal coupling field $\boldsymbol{E_2}$. Under the EIT condition, the susceptibility pattern can form a honeycomb structure with the lattice constant being 76 $\mu$m. There exists a frequency difference of 3.03~GHz between the probe and coupling fields. The powers of the three coupling beams are 25 mW, 20 mW and 17 mW. ECDL3 emits two pump beams (wavelength $\approx$ 780.2 nm, with power being 12.5 mW and 6.5 mW) and they interfere to establish a one-dimensional pump field (vertically polarized) to cover only A sites. The $\mu$-metal-wrapped Rb vapor cell is 5~cm in length and is heated to $110^{\circ}$C by a home-made temperature controller. A polarization beam splitter is placed behind the output surface of the cell to prevent the coupling and pump beams entering the camera.

We analyze the output intensity distribution captured by the CCD camera with a resolution of $1562\times 611$ pixels. The experimental uncertainty on the wave packet width and center of mass position marked as error bars in Figs.~\ref{fig2},\ref{fig3} of the main text originates mostly from the cutoff used to remove the noise in the regions without any signal. The choice of the corresponding threshold affects the spatial averaging. The uncertainty for the non-Hermiticity is determined from the calibration curve using the model discussed below.

\noindent{\textbf{Theoretical model for the susceptibility.}} We consider a four-level $N$-type $^{85}Rb$ atomic configuration given in Extended Data Fig.~\ref{figS1}, in which the probe $\textit{\textbf{E}}_1$, coupling $\textit{\textbf{E}}_2$, and pump $\textit{\textbf{E}}_3$ fields drive transitions $\left |1\right \rangle$ $\xrightarrow{}$ $\left |3\right \rangle$, $\left |2\right \rangle$ $\xrightarrow{}$ $\left |3\right \rangle$ and $\left |1\right \rangle$ $\xrightarrow{}$ $\left |4\right \rangle$ respectively. Specifically, the pump field $\textit{\textbf{E}}_3$ arising from two-beam interference covers one set of sublattices in the honeycomb photonic lattice (formed by the coupling field $\textit{\textbf{E}}_2$ with a hexagonal intensity distribution) to adjust the imaginary part of the susceptibility on the sites. The interaction between laser beams and the atomic system under the rotation wave approximation can be described by the density matrix, and the density matrix equations can be expressed in the following form~\cite{PhysRevA.88.041803}:
\begin{eqnarray}
\label{eqrho}
     \Dot{\rho}_{22}&=&\Gamma_{42}\rho_{44}+\Gamma_{32}\rho_{33}-\Gamma_{21}\rho_{22}+\frac{i}{2}(\rho_{32}-\rho_{23})\Omega_2,  \\
     \Dot{\rho}_{33}&=&\Gamma_{43}\rho_{44}-\Gamma_{32}\rho_{33}-\Gamma_{31}\rho_{33}+\frac{i}{2}[(\rho_{23}-\rho_{32}) \Omega_2+(\rho_{13}-\rho_{31}) \Omega_1],\nonumber \\
     \Dot{\rho}_{44}&=&-(\Gamma_{43}+\Gamma_{42}+\Gamma_{41})\rho_{44}+\frac{i}{2}(\rho_{14}-\rho_{41}) \Omega_3, \nonumber\\
     \Dot{\rho}_{21}&=&-\tilde{\gamma}_{21}\rho_{21}+\frac{i}{2}(\rho_{31}\Omega_2-\rho_{24}\Omega_3-\rho_{23}\Omega_1),\nonumber\\
     \Dot{\rho}_{31}&=&-\tilde{\gamma}_{31}\rho_{31}+\frac{i}{2}[\rho_{21}\Omega_2-\rho_{34}\Omega_3+(\rho_{11}-\rho_{33})\Omega_1],\nonumber \\
     \Dot{\rho}_{41}&=&-\tilde{\gamma}_{41}\rho_{41}+\frac{i}{2}[-\rho_{43}\Omega_1+(\rho_{11}-\rho_{44})\Omega_3], \nonumber\\
     \Dot{\rho}_{32}&=&-\tilde{\gamma}_{32}\rho_{32}+\frac{i}{2}[\rho_{12}\Omega_1+(\rho_{22}-\rho_{33})\Omega_2], \nonumber\\
     \Dot{\rho}_{42}&=&-\tilde{\gamma}_{42}\rho_{42}+\frac{i}{2}(\rho_{12}\Omega_3-\rho_{43}\Omega_2),\nonumber \\
     \Dot{\rho}_{43}&=&-\tilde{\gamma}_{43}\rho_{43}+\frac{i}{2}(\rho_{13}\Omega_3-\rho_{42}\Omega_2-\rho_{41}\Omega_1),\nonumber
\end{eqnarray}
and 
\begin{equation}
     \rho_{11}+\rho_{22}+\rho_{33}+\rho_{44}=1.
\end{equation}
Here, $\Gamma_{mn}$ is the natural decay rate between states $\left |m\right \rangle$ and $\left |n\right \rangle$. Other parameters are defined as: $ \tilde{\gamma}_{21}=\gamma_{21}-i(\Delta_1-\Delta_2)$, $\tilde{\gamma}_{31}=\gamma_{31}-i\Delta_1$, $\tilde{\gamma}_{32}=\gamma_{32}-i\Delta_2$ , $\tilde{\gamma}_{41}=\gamma_{41}-i\Delta_3$, $\tilde{\gamma}_{42}=\gamma_{42}-i(\Delta_2+\Delta_3-\Delta_1)$, $\tilde{\gamma}_{43}=\gamma_{43}-i(\Delta_3-\Delta_1)$ with $\gamma_{mn}=(\Gamma_m+\Gamma_n)/2$ and $\Delta_i$  $(i=1,2,3)$ being the frequency detuning. $\Omega_i$ is the Rabi freqquency of field $\textit{\textbf{E}}_i$, and directly proportional to its electric intensity $E_i$. Terms $\rho_{mn}$ are the density-matrix elements corresponding to the transition from $\left |m\right \rangle$ to $\left |n\right \rangle$ ($m, n=1, 2, 3$ and $4$). According to Eq.~\eqref{eqrho} one can obtain $\rho_{31}$ describing the response of the multi-level atomic system to the probe beam, and the spatial distribution of susceptibility can be expressed as:
 \begin{equation}
   \chi=\chi_r+\chi_i=\frac{2N\mu_{13}\rho_{31}}{\epsilon_0 \textit{E}_1}
\end{equation}
Parameters in this formula are given as: $N$ represents atomic density, $\epsilon_0$ represents vacuum dielectric constant, $E_1$ is the electric field strength of the probe beam and $\mu_{13}$ is dipole momentum between $\left |1\right \rangle$ and $\left |3\right \rangle$.
By setting the two-photon detuning $\Delta_1 - \Delta_2 >0$, a non-Hermitian honeycomb photonic lattice is constructed, and the pump field can modify the difference in imaginary parts $\chi_i$ between the two types of sublattice (A and B)~\cite{Feng2023}. 
When certain specific parameters are selected, the pump field only affects the distribution of the imaginary part $\chi_i$ but hardly affects the real part $\chi_r$. In Extended Data Fig.~\ref{figS2}, the distributions of the real and imaginary parts of the susceptibility under a certain set of parameters is plotted. In panel a, it can be seen that the real part still presents a uniform honeycomb profile (its staggering is negligible), while panel b shows that the imaginary parts of A and B sublattices are remarkably different. The adopted parameters are: $\Omega_1=0.259\pi$ MHz, $\Omega_2=120\pi$ MHz, $\Omega_3=2.5\pi$ MHz, $\Delta_1=-50$ MHz, $\Delta_2=-80$ MHz, and $\Delta_3=40$ MHz. The difference of $\chi_i$ in the A and B lattice sites can be easily controlled by adjusting parameters such as detuning and intensity of the pump field. We are using the peak values of the imaginary part of the susceptibility for $\Gamma_A$ and $\Gamma_B$ in the main text.

\noindent{\textbf{Theoretical description of the exceptional ring.}}
We begin with the right eigenstates of the Hamiltonian in Eq.~\eqref{HamNH}, which are given by $\ket{\psi_+}=(ke^{-i\theta}/\sqrt{2\gamma(\gamma+\sqrt{\gamma^2-k^2})},k/\sqrt{2\gamma(\gamma-\sqrt{\gamma^2-k^2})})^T$ and $\ket{\psi_-}=(\sqrt{\gamma+\sqrt{\gamma^2-k^2}}e^{-i\theta}/\sqrt{2\gamma},k/\sqrt{2\gamma(\gamma+\sqrt{\gamma^2-k^2})})^T$ (inside the ring) and by $\ket{\psi_+}=((i\gamma-\sqrt{k^2-\gamma^2}e^{-i\theta})/k\sqrt{2}, 1/\sqrt{2})^T$ and $\ket{\psi_-}=(i\gamma+\sqrt{k^2-\gamma^2}e^{-i\theta})/k\sqrt{2}, 1/\sqrt{2})^T$ (outside the ring). These eigenstates are normalized, allowing to calculate the pseudospin projections shown in Extended Data Fig.~\ref{EDFpseudoNOmetric}(a,b). The pseudospin representation~\cite{feynman1957geometrical} allows to understand the qualitative behavior of these eigenstates: panel a shows the pseudospin (whose components are defined as the average values of the spin operators, $S_i=\bra{\psi}\sigma_i\ket{\psi}$) as a function of a 2D wave vector $k_x$, $k_y$ (arrows represent the in-plane projection $(S_x,S_y)$ and the false color represents $S_z$) and panel b shows the three pseudospin components (black, red, and blue) as a function of $k_x$ for $k_y=0$. Finally, the non-orthogonality of the eigenstates (the overlap between the two right eigenvectors $\ket{\psi_+}$ and $\ket{\psi_-}$) is also shown as a magenta curve.  It decreases relatively slowly with the wave vector outside the ring (it is still 10\% for $k=10 \gamma$).

For the calculation of the quantum metric we use a different normalization condition, as stated in the main text: $\bra{\phi_\pm}\ket{\psi_\pm}=1$. As an example, we provide here two such normalized eigenvectors (left and right):
\begin{equation}
\left\langle {{\phi _ + }} \right| = \bra{
{ - \frac{{\gamma  - \sqrt {{\gamma ^2} - {k^2}} }}{{\sqrt {2\left( {{k^2} - \gamma \left( {\gamma  - \sqrt {{\gamma ^2} - {k^2}} } \right)} \right)} }}{e^{  i\theta }}},\,
{\frac{k}{{\sqrt {2\left( {{k^2} - \gamma \left( {\gamma  - \sqrt {{\gamma ^2} - {k^2}} } \right)} \right)} }}}
 }
\end{equation}
and
\begin{equation}
\left| {{\psi _ + }} \right\rangle  = \left| {\begin{array}{*{20}{c}}
{\frac{{\gamma  - \sqrt {{\gamma ^2} - {k^2}} }}{{\sqrt {2\left( {{k^2} - \gamma \left( {\gamma  - \sqrt {{\gamma ^2} - {k^2}} } \right)} \right)} }}{e^{ - i\theta }}}\\
{\frac{k}{{\sqrt {2\left( {{k^2} - \gamma \left( {\gamma  - \sqrt {{\gamma ^2} - {k^2}} } \right)} \right)} }}}
\end{array}} \right\rangle 
\end{equation}

Using the definition of the biorthogonal quantum metric tensor in Eq.~\eqref{QMbiorth} we obtain the following expressions. Inside the ring, $g_{kk}\approx (5k^2+2\gamma^2)/8\gamma^4$ , $g_{\theta\theta}\approx k^2/4\gamma^2$ (the full expressions are really cumbersome). Outside of the ring, $g_{kk}=\gamma^4/4(k^2-\gamma^2)^2$, $g_{\theta\theta}=k^2/4(k^2-\gamma^2)$. The standard quantum metric reads $g^{Herm}_{kk}=1/4(\gamma^2-k^2)$, $g^{Herm}_{\theta\theta}=k^2/4\gamma^2$ inside the ring and $g^{Herm}_{kk}=\gamma^2/4k^2(k^2-\gamma^2)$, $g^{Herm}_{\theta\theta}=1/4$ outside the ring.

The components of the biorthogonal quantum metric coincide with the standard quantum metric in the limit of zero non-orthogonality ($k\to 0$ and $k\to\infty$), but exhibit an important difference close to the exceptional ring $k=\gamma$. In particular, the azimuthal component of the biorthogonal quantum metric diverges there, whereas the standard one does not. This difference shows up in the transverse non-Hermitian drift, as discussed in the main text.
All these results are summarized in Extended Data Fig.~\ref{EDFpseudoNOmetric}c,d. The biorthogonal quantum metric is shown with black curves, and the standard one with blue curves.

The calculation of the overlap with the quantum metric involves integration in the parameter space, with the limits of the integration defined by the two wave functions involved. For the real-space broadening presented in Fig.~\ref{fig2}, we take the initial spinor $\ket{\psi(0)}=(1,0)^T$, corresponding to the parameter space limit $\bm{k}=0$. In the experiment, it may be different from this perfect spinor, which shows up as a slight deviation from a perfect ring in Fig.~\ref{fig2}a (Hermitian limit). However, for large values of non-Hermiticity this small deviation does not play any role because of the fast decay of the second component. For the reciprocal-space center of mass dynamics, we take the initial spinor $\ket{\psi(0)=(1,1)^T/\sqrt{2}}$, corresponding to the excitation of A and B lattices with equal amplitude. This is also what the experiment was targetting at. Small deviations from this spinor can induce a non-trivial real-space dynamics (e.g. Zitterbewegung), but not a reciprocal space one. This spinor corresponds to the parameter space point $\bm{k}=(+\infty,0)$, which is what we use in the main text.

\noindent{\textbf{Paraxial approximation.}}
The propagation of a probe beam through the four-level atomic vapors with an EIT-induced spatially periodic susceptibility distribution is described by the paraxial equation:
\begin{equation}
    i\frac{\partial E}{\partial z}=-\frac{1}{2k_0}\Delta E-\frac{k_0\chi}{2}E\text{,}
    \label{paraxial}
\end{equation}
where $k_0$ is the probe wave vector. This is equivalent to a 2D time-dependent Schr\"odinger equation. To solve the paraxial equation~\eqref{paraxial}, we used the combination of the 3rd-order Adams-Bashforth method with Fourier-transform calculation of the Laplacian operator accelerated by the Graphics Processor Unit. The real and imaginary parts of the susceptibility were calculated as discussed above. The dispersions shown in Fig.~\ref{fig2} were calculated using a narrow probe excitation (smaller than a single lattice site). The resulting solution $E(x,y,z)$ was Fourier-transformed over all coordinates to obtain the dispersion $|E(k_x,k_y,\hbar\omega)|^2$ (with $\hbar\omega=\hbar c k_z$), whose cuts are shown in Fig.~\ref{fig2}c,d. 
The probe was introduced as an initial condition ($z=0$) for the complex electric field $E$. 

\noindent{\textbf{Analytical solution at short times and low non-Hermiticity.}} The analytical solution at short times shown in Fig.~\ref{fig2}e as a red dashed line is governed by the dominating Hermitian terms of the Hamiltonian and thus does not explicitly involve the biorthogonal quantum metric (it is still possible to rewrite it using the standard quantum metric), but we provide this solution here in brief, for completeness. To obtain this solution, we replace the wave vectors in the Hamiltonian in Eq.~\eqref{HamNH} by operators $\bm{k}=-i\nabla$. These terms provide a coupling between components. If one component is excited with a Gaussian $G_1(r)$, the second component is populated by $-i\nabla G_1(r)$, and the first component is in turn re-populated by $(-i\nabla)^2G_1(r)$. At the same time, the components exhibit amplification and decay by $\exp(\pm\gamma t)$ (neglecting the average losses). Calculating the root mean square width as $\sigma=\sqrt{\int\limits_0^\infty r^2 p(r) 2\pi r\, dr}$ allows obtaining the following expression
\begin{equation}
    \sigma\approx\sqrt{\frac{3\sigma_0^2t^4+4\sigma_0^4t^2 e^{2\gamma t}+\sigma_0^6 e^{4\gamma t}}{t^4+\sigma_0^2t^2 e^{2\gamma t}+\sigma_0^4 e^{4\gamma t}}}
\end{equation}
Here, the dynamics is mostly Hermitian, since the wave packet is mostly located outside the ER. While it can be also described with the quantum metric~\cite{leblanc2021universal}, in the present manuscript we focus on the use of the biorthogonal quantum  metric for the strongly non-Hermitian case.

In experiments, we use this expression for fixed $t$, corresponding to the vapor cell length, and vary the non-Hermiticity $\gamma$. The red dashed line obtained with the expression above is in a very good agreement ($R^2=0.943$) with the experimental data for low non-Hermiticity.

\end{methods}

\section*{Data availability}
The data generated in this study are available in the Open Science Framework (OSF) repository:
\verb|https://osf.io/xpnkh/?view_only=03e5151e6690438fa65f0711cab9d944|

\section*{Authors contribution}
Using the CRediT contributor roles taxonomy. 
Conceptualisation: IS, DS, PK, GM, ZZ, LZ.
Data curation: ZZ, LS, DS.
Formal Analysis: IS,  DS, ZZ, LZ.
Funding acquisition: DS, GM, ZZ, LM.
Investigation: IS, DS, ZZ, LS, LC.
Methodology: IS, DS, ZZ, XM, ZY, WH.
Project administration: IS, DS, ZZ.
Software: DS, LZ, LF.
Supervision: IS, DS, GM, ZZ, XM.
Validation: IS, DS, PK, ZZ.
Visualization: IS, DS, ZZ.
Writing – original draft: IS, DS, ZZ.
Writing – review \& editing: IS, PK, DS, ZZ, LZ, LC.

\section*{References}
\bibliographystyle{naturemag}
\bibliography{biblio}

\begin{thebibliography}{10}
\expandafter\ifx\csname url\endcsname\relax
  \def\url#1{\texttt{#1}}\fi
\expandafter\ifx\csname urlprefix\endcsname\relax\def\urlprefix{URL }\fi
\providecommand{\bibinfo}[2]{#2}
\providecommand{\eprint}[2][]{\url{#2}}

\bibitem{Moiseyev2011}
\bibinfo{author}{Moiseyev, N.}
\newblock \emph{\bibinfo{title}{Non-Hermitian quantum mechanics}} (\bibinfo{publisher}{Cambridge University Press, UK}, \bibinfo{year}{2011}).

\bibitem{Miri2019}
\bibinfo{author}{Miri, M.~A.} \& \bibinfo{author}{Al{\`{u}}, A.}
\newblock \bibinfo{title}{{Exceptional points in optics and photonics}}.
\newblock \emph{\bibinfo{journal}{Science}} \textbf{\bibinfo{volume}{363}} (\bibinfo{year}{2019}).

\bibitem{Wiersig2014}
\bibinfo{author}{Wiersig, J.}
\newblock \bibinfo{title}{Enhancing the sensitivity of frequency and energy splitting detection by using exceptional points: Application to microcavity sensors for single-particle detection}.
\newblock \emph{\bibinfo{journal}{Phys. Rev. Lett.}} \textbf{\bibinfo{volume}{112}}, \bibinfo{pages}{203901} (\bibinfo{year}{2014}).

\bibitem{Wiersig2016}
\bibinfo{author}{Wiersig, J.}
\newblock \bibinfo{title}{{Sensors operating at exceptional points: General theory}}.
\newblock \emph{\bibinfo{journal}{Phys. Rev. A}} \textbf{\bibinfo{volume}{93}}, \bibinfo{pages}{1--9} (\bibinfo{year}{2016}).

\bibitem{hodaei2017enhanced}
\bibinfo{author}{Hodaei, H.} \emph{et~al.}
\newblock \bibinfo{title}{Enhanced sensitivity at higher-order exceptional points}.
\newblock \emph{\bibinfo{journal}{Nature}} \textbf{\bibinfo{volume}{548}}, \bibinfo{pages}{187--191} (\bibinfo{year}{2017}).

\bibitem{chen2017exceptional}
\bibinfo{author}{Chen, W.}, \bibinfo{author}{{\"O}zdemir, {\c{S}}.~K.}, \bibinfo{author}{Zhao, G.}, \bibinfo{author}{Wiersig, J.} \& \bibinfo{author}{Yang, L.}
\newblock \bibinfo{title}{Exceptional points enhance sensing in an optical microcavity}.
\newblock \emph{\bibinfo{journal}{Nature}} \textbf{\bibinfo{volume}{548}}, \bibinfo{pages}{192--196} (\bibinfo{year}{2017}).

\bibitem{park2020symmetry}
\bibinfo{author}{Park, J.-H.} \emph{et~al.}
\newblock \bibinfo{title}{Symmetry-breaking-induced plasmonic exceptional points and nanoscale sensing}.
\newblock \emph{\bibinfo{journal}{Nat. Phys.}} \textbf{\bibinfo{volume}{16}}, \bibinfo{pages}{462--468} (\bibinfo{year}{2020}).

\bibitem{duggan2022limitations}
\bibinfo{author}{Duggan, R.}, \bibinfo{author}{Mann, S.~A.} \& \bibinfo{author}{Al\`u, A.}
\newblock \bibinfo{title}{Limitations of sensing at an exceptional point}.
\newblock \emph{\bibinfo{journal}{ACS Photonics}} \textbf{\bibinfo{volume}{9}}, \bibinfo{pages}{1554--1566} (\bibinfo{year}{2022}).

\bibitem{Wiersig2022}
\bibinfo{author}{Wiersig, J.}
\newblock \bibinfo{title}{Distance between exceptional points and diabolic points and its implication for the response strength of non-hermitian systems}.
\newblock \emph{\bibinfo{journal}{Phys. Rev. Res.}} \textbf{\bibinfo{volume}{4}}, \bibinfo{pages}{033179} (\bibinfo{year}{2022}).

\bibitem{Liao2021}
\bibinfo{author}{Liao, Q.} \emph{et~al.}
\newblock \bibinfo{title}{Experimental measurement of the divergent quantum metric of an exceptional point}.
\newblock \emph{\bibinfo{journal}{Phys. Rev. Lett.}} \textbf{\bibinfo{volume}{127}}, \bibinfo{pages}{107402} (\bibinfo{year}{2021}).

\bibitem{Zanardi2007}
\bibinfo{author}{Zanardi, P.}, \bibinfo{author}{Giorda, P.} \& \bibinfo{author}{Cozzini, M.}
\newblock \bibinfo{title}{Information-theoretic differential geometry of quantum phase transitions}.
\newblock \emph{\bibinfo{journal}{Phys. Rev. Lett.}} \textbf{\bibinfo{volume}{99}}, \bibinfo{pages}{100603} (\bibinfo{year}{2007}).

\bibitem{zhen2015spawning}
\bibinfo{author}{Zhen, B.} \emph{et~al.}
\newblock \bibinfo{title}{Spawning rings of exceptional points out of {D}irac cones}.
\newblock \emph{\bibinfo{journal}{Nature}} \textbf{\bibinfo{volume}{525}}, \bibinfo{pages}{354--358} (\bibinfo{year}{2015}).

\bibitem{xu2017weyl}
\bibinfo{author}{Xu, Y.}, \bibinfo{author}{Wang, S.-T.} \& \bibinfo{author}{Duan, L.-M.}
\newblock \bibinfo{title}{Weyl exceptional rings in a three-dimensional dissipative cold atomic gas}.
\newblock \emph{\bibinfo{journal}{Phys. Rev. Lett.}} \textbf{\bibinfo{volume}{118}}, \bibinfo{pages}{045701} (\bibinfo{year}{2017}).

\bibitem{cerjan2019experimental}
\bibinfo{author}{Cerjan, A.} \emph{et~al.}
\newblock \bibinfo{title}{Experimental realization of a {W}eyl exceptional ring}.
\newblock \emph{\bibinfo{journal}{Nat. Photonics}} \textbf{\bibinfo{volume}{13}}, \bibinfo{pages}{623--628} (\bibinfo{year}{2019}).

\bibitem{yoshida2019exceptional}
\bibinfo{author}{Yoshida, T.} \& \bibinfo{author}{Hatsugai, Y.}
\newblock \bibinfo{title}{Exceptional rings protected by emergent symmetry for mechanical systems}.
\newblock \emph{\bibinfo{journal}{Phys. Rev. B}} \textbf{\bibinfo{volume}{100}}, \bibinfo{pages}{054109} (\bibinfo{year}{2019}).

\bibitem{xu2022observation}
\bibinfo{author}{Xu, G.} \emph{et~al.}
\newblock \bibinfo{title}{Observation of weyl exceptional rings in thermal diffusion}.
\newblock \emph{\bibinfo{journal}{Proceedings of the National Academy of Sciences (PNAS)}} \textbf{\bibinfo{volume}{119}}, \bibinfo{pages}{e2110018119} (\bibinfo{year}{2022}).

\bibitem{liu2022experimental}
\bibinfo{author}{Liu, J.-j.} \emph{et~al.}
\newblock \bibinfo{title}{Experimental realization of weyl exceptional rings in a synthetic three-dimensional non-hermitian phononic crystal}.
\newblock \emph{\bibinfo{journal}{Phys. Rev. Lett.}} \textbf{\bibinfo{volume}{129}}, \bibinfo{pages}{084301} (\bibinfo{year}{2022}).

\bibitem{li2023exceptional}
\bibinfo{author}{Li, A.} \emph{et~al.}
\newblock \bibinfo{title}{Exceptional points and non-hermitian photonics at the nanoscale}.
\newblock \emph{\bibinfo{journal}{Nat. Nanotechnol.}} \textbf{\bibinfo{volume}{18}}, \bibinfo{pages}{706--720} (\bibinfo{year}{2023}).

\bibitem{Leclerc2024}
\bibinfo{author}{Leclerc, A.} \emph{et~al.}
\newblock \bibinfo{title}{Exceptional ring of the buoyancy instability in stars}.
\newblock \emph{\bibinfo{journal}{Phys. Rev. Res.}} \textbf{\bibinfo{volume}{6}}, \bibinfo{pages}{L012055} (\bibinfo{year}{2024}).

\bibitem{Sundaram1999}
\bibinfo{author}{Sundaram, G.} \& \bibinfo{author}{Niu, Q.}
\newblock \bibinfo{title}{Wave-packet dynamics in slowly perturbed crystals: Gradient corrections and berry-phase effects}.
\newblock \emph{\bibinfo{journal}{Phys. Rev. B}} \textbf{\bibinfo{volume}{59}}, \bibinfo{pages}{14915--14925} (\bibinfo{year}{1999}).

\bibitem{Bliokh2007}
\bibinfo{author}{Bliokh, K.~Y.}, \bibinfo{author}{Bliokh, Y.~P.}, \bibinfo{author}{Savel'ev, S.} \& \bibinfo{author}{Nori, F.}
\newblock \bibinfo{title}{Semiclassical dynamics of electron wave packet states with phase vortices}.
\newblock \emph{\bibinfo{journal}{Phys. Rev. Lett.}} \textbf{\bibinfo{volume}{99}}, \bibinfo{pages}{190404} (\bibinfo{year}{2007}).

\bibitem{Torma2018}
\bibinfo{author}{T\"orm\"a, P.}, \bibinfo{author}{Liang, L.} \& \bibinfo{author}{Peotta, S.}
\newblock \bibinfo{title}{Quantum metric and effective mass of a two-body bound state in a flat band}.
\newblock \emph{\bibinfo{journal}{Phys. Rev. B}} \textbf{\bibinfo{volume}{98}}, \bibinfo{pages}{220511} (\bibinfo{year}{2018}).

\bibitem{Bleu2018effective}
\bibinfo{author}{Bleu, O.}, \bibinfo{author}{Malpuech, G.}, \bibinfo{author}{Gao, Y.} \& \bibinfo{author}{Solnyshkov, D.~D.}
\newblock \bibinfo{title}{Effective theory of nonadiabatic quantum evolution based on the quantum geometric tensor}.
\newblock \emph{\bibinfo{journal}{Phys. Rev. Lett.}} \textbf{\bibinfo{volume}{121}}, \bibinfo{pages}{020401} (\bibinfo{year}{2018}).

\bibitem{leblanc2021universal}
\bibinfo{author}{Leblanc, C.}, \bibinfo{author}{Malpuech, G.} \& \bibinfo{author}{Solnyshkov, D.}
\newblock \bibinfo{title}{Universal semiclassical equations based on the quantum metric for a two-band system}.
\newblock \emph{\bibinfo{journal}{Phys. Rev. B}} \textbf{\bibinfo{volume}{104}}, \bibinfo{pages}{134312} (\bibinfo{year}{2021}).

\bibitem{Torma2023essay}
\bibinfo{author}{T\"orm\"a, P.}
\newblock \bibinfo{title}{Essay: Where can quantum geometry lead us?}
\newblock \emph{\bibinfo{journal}{Phys. Rev. Lett.}} \textbf{\bibinfo{volume}{131}}, \bibinfo{pages}{240001} (\bibinfo{year}{2023}).

\bibitem{hu2024generalized}
\bibinfo{author}{Hu, Y.-M.~R.}, \bibinfo{author}{Ostrovskaya, E.~A.} \& \bibinfo{author}{Estrecho, E.}
\newblock \bibinfo{title}{Generalized quantum geometric tensor in a non-{H}ermitian exciton-polariton system}.
\newblock \emph{\bibinfo{journal}{Opt. Mater. Express}} \textbf{\bibinfo{volume}{14}}, \bibinfo{pages}{664--686} (\bibinfo{year}{2024}).

\bibitem{provost1980riemannian}
\bibinfo{author}{Provost, J.} \& \bibinfo{author}{Vallee, G.}
\newblock \bibinfo{title}{Riemannian structure on manifolds of quantum states}.
\newblock \emph{\bibinfo{journal}{Commun. Math. Phys.}} \textbf{\bibinfo{volume}{76}}, \bibinfo{pages}{289--301} (\bibinfo{year}{1980}).

\bibitem{Brody2013}
\bibinfo{author}{Brody, D.~C.} \& \bibinfo{author}{Graefe, E.~M.}
\newblock \bibinfo{title}{{Information geometry of complex hamiltonians and exceptional points}}.
\newblock \emph{\bibinfo{journal}{Entropy}} \textbf{\bibinfo{volume}{15}}, \bibinfo{pages}{3361--3378} (\bibinfo{year}{2013}).

\bibitem{brody2013biorthogonal}
\bibinfo{author}{Brody, D.~C.}
\newblock \bibinfo{title}{Biorthogonal quantum mechanics}.
\newblock \emph{\bibinfo{journal}{J. Phys. A-Math. Theor.}} \textbf{\bibinfo{volume}{47}}, \bibinfo{pages}{035305} (\bibinfo{year}{2013}).

\bibitem{moiseyev1978resonance}
\bibinfo{author}{Moiseyev, N.}, \bibinfo{author}{Certain, P.} \& \bibinfo{author}{Weinhold, F.}
\newblock \bibinfo{title}{Resonance properties of complex-rotated hamiltonians}.
\newblock \emph{\bibinfo{journal}{Mol. Phys.}} \textbf{\bibinfo{volume}{36}}, \bibinfo{pages}{1613--1630} (\bibinfo{year}{1978}).

\bibitem{solnyshkov2021quantum}
\bibinfo{author}{Solnyshkov, D.} \emph{et~al.}
\newblock \bibinfo{title}{Quantum metric and wave packets at exceptional points in non-hermitian systems}.
\newblock \emph{\bibinfo{journal}{Phys. Rev. B}} \textbf{\bibinfo{volume}{103}}, \bibinfo{pages}{125302} (\bibinfo{year}{2021}).

\bibitem{Hu2023}
\bibinfo{author}{Hu, Y.-M.~R.}, \bibinfo{author}{Ostrovskaya, E.~A.} \& \bibinfo{author}{Estrecho, E.}
\newblock \bibinfo{title}{Wave-packet dynamics in a non-hermitian exciton-polariton system}.
\newblock \emph{\bibinfo{journal}{Phys. Rev. B}} \textbf{\bibinfo{volume}{108}}, \bibinfo{pages}{115404} (\bibinfo{year}{2023}).

\bibitem{Ye2024}
\bibinfo{author}{Chen~Ye, C.}, \bibinfo{author}{Vleeshouwers, W.~L.}, \bibinfo{author}{Heatley, S.}, \bibinfo{author}{Gritsev, V.} \& \bibinfo{author}{Morais~Smith, C.}
\newblock \bibinfo{title}{Quantum metric of non-hermitian su-schrieffer-heeger systems}.
\newblock \emph{\bibinfo{journal}{Phys. Rev. Res.}} \textbf{\bibinfo{volume}{6}}, \bibinfo{pages}{023202} (\bibinfo{year}{2024}).

\bibitem{gianfrate2020measurement}
\bibinfo{author}{Gianfrate, A.} \emph{et~al.}
\newblock \bibinfo{title}{Measurement of the quantum geometric tensor and of the anomalous {H}all drift}.
\newblock \emph{\bibinfo{journal}{Nature}} \textbf{\bibinfo{volume}{578}}, \bibinfo{pages}{381--385} (\bibinfo{year}{2020}).

\bibitem{Yu2019}
\bibinfo{author}{Yu, M.} \emph{et~al.}
\newblock \bibinfo{title}{{Experimental measurement of the quantum geometric tensor using coupled qubits in diamond}}.
\newblock \emph{\bibinfo{journal}{Natl. Sci. Rev.}} \textbf{\bibinfo{volume}{7}}, \bibinfo{pages}{254--260} (\bibinfo{year}{2019}).

\bibitem{Cuerda2024}
\bibinfo{author}{Cuerda, J.}, \bibinfo{author}{Taskinen, J.~M.}, \bibinfo{author}{K\"allman, N.}, \bibinfo{author}{Grabitz, L.} \& \bibinfo{author}{T\"orm\"a, P.}
\newblock \bibinfo{title}{Observation of quantum metric and non-hermitian berry curvature in a plasmonic lattice}.
\newblock \emph{\bibinfo{journal}{Phys. Rev. Res.}} \textbf{\bibinfo{volume}{6}}, \bibinfo{pages}{L022020} (\bibinfo{year}{2024}).

\bibitem{PhysRevLett.74.666}
\bibinfo{author}{Xiao, M.}, \bibinfo{author}{Li, Y.-q.}, \bibinfo{author}{Jin, S.-z.} \& \bibinfo{author}{Gea-Banacloche, J.}
\newblock \bibinfo{title}{Measurement of dispersive properties of electromagnetically induced transparency in rubidium atoms}.
\newblock \emph{\bibinfo{journal}{Phys. Rev. Lett.}} \textbf{\bibinfo{volume}{74}}, \bibinfo{pages}{666--669} (\bibinfo{year}{1995}).

\bibitem{Feng2023}
\bibinfo{author}{Feng, Y.} \emph{et~al.}
\newblock \bibinfo{title}{Loss difference induced localization in a non-hermitian honeycomb photonic lattice}.
\newblock \emph{\bibinfo{journal}{Phys. Rev. Lett.}} \textbf{\bibinfo{volume}{131}}, \bibinfo{pages}{013802} (\bibinfo{year}{2023}).

\bibitem{zhao2019non}
\bibinfo{author}{Zhao, H.} \emph{et~al.}
\newblock \bibinfo{title}{Non-hermitian topological light steering}.
\newblock \emph{\bibinfo{journal}{Science}} \textbf{\bibinfo{volume}{365}}, \bibinfo{pages}{1163--1166} (\bibinfo{year}{2019}).

\bibitem{heck2017highly}
\bibinfo{author}{Heck, M.~J.}
\newblock \bibinfo{title}{Highly integrated optical phased arrays: photonic integrated circuits for optical beam shaping and beam steering}.
\newblock \emph{\bibinfo{journal}{Nanophotonics}} \textbf{\bibinfo{volume}{6}}, \bibinfo{pages}{93--107} (\bibinfo{year}{2017}).

\bibitem{PhysRevLett.129.233901}
\bibinfo{author}{Zhang, Z.} \emph{et~al.}
\newblock \bibinfo{title}{Angular-dependent klein tunneling in photonic graphene}.
\newblock \emph{\bibinfo{journal}{Phys. Rev. Lett.}} \textbf{\bibinfo{volume}{129}}, \bibinfo{pages}{233901} (\bibinfo{year}{2022}).

\bibitem{zhang2020observation}
\bibinfo{author}{Zhang, Z.} \emph{et~al.}
\newblock \bibinfo{title}{Observation of edge solitons in photonic graphene}.
\newblock \emph{\bibinfo{journal}{Nat. Commun.}} \textbf{\bibinfo{volume}{11}}, \bibinfo{pages}{1902} (\bibinfo{year}{2020}).

\bibitem{PhysRevLett.132.263801}
\bibinfo{author}{Zhang, Z.} \emph{et~al.}
\newblock \bibinfo{title}{Non-hermitian delocalization in a two-dimensional photonic quasicrystal}.
\newblock \emph{\bibinfo{journal}{Phys. Rev. Lett.}} \textbf{\bibinfo{volume}{132}}, \bibinfo{pages}{263801} (\bibinfo{year}{2024}).

\bibitem{Hamilton1837}
\bibinfo{author}{Hamilton, W.~R.}
\newblock \bibinfo{title}{Third supplement to an essay on the theory of systems of rays}.
\newblock \emph{\bibinfo{journal}{Trans. Royal Irish Acad.}} \textbf{\bibinfo{volume}{17}}, \bibinfo{pages}{1--144} (\bibinfo{year}{1837}).

\bibitem{Lloyd1837}
\bibinfo{author}{Lloyd, H.}
\newblock \bibinfo{title}{On the phenomena presented by light in its passage along the axes of biaxial crystals}.
\newblock \emph{\bibinfo{journal}{Trans. Roayl Irish Acad.}} \textbf{\bibinfo{volume}{17}}, \bibinfo{pages}{145} (\bibinfo{year}{1837}).

\bibitem{berry2006conical}
\bibinfo{author}{Berry, M.}, \bibinfo{author}{Jeffrey, M.} \& \bibinfo{author}{Lunney, J.}
\newblock \bibinfo{title}{Conical diffraction: observations and theory}.
\newblock \emph{\bibinfo{journal}{Proc. R. Soc. A}} \textbf{\bibinfo{volume}{462}}, \bibinfo{pages}{1629--1642} (\bibinfo{year}{2006}).

\bibitem{Zhang2019}
\bibinfo{author}{Zhang, Z.} \emph{et~al.}
\newblock \bibinfo{title}{Particlelike behavior of topological defects in linear wave packets in photonic graphene}.
\newblock \emph{\bibinfo{journal}{Phys. Rev. Lett.}} \textbf{\bibinfo{volume}{122}}, \bibinfo{pages}{233905} (\bibinfo{year}{2019}).

\bibitem{ouyang2007light}
\bibinfo{author}{Ouyang, S.}, \bibinfo{author}{Hu, W.} \& \bibinfo{author}{Guo, Q.}
\newblock \bibinfo{title}{Light steering in a strongly nonlocal nonlinear medium}.
\newblock \emph{\bibinfo{journal}{Phys. Rev. A}} \textbf{\bibinfo{volume}{76}}, \bibinfo{pages}{053832} (\bibinfo{year}{2007}).

\bibitem{Luk2021}
\bibinfo{author}{Luk, S. M.~H.} \emph{et~al.}
\newblock \bibinfo{title}{All-optical beam steering using the polariton lighthouse effect}.
\newblock \emph{\bibinfo{journal}{ACS Photonics}} \textbf{\bibinfo{volume}{8}}, \bibinfo{pages}{449--454} (\bibinfo{year}{2021}).
\newblock \urlprefix\url{https://doi.org/10.1021/acsphotonics.0c01962}.
\newblock \eprint{https://doi.org/10.1021/acsphotonics.0c01962}.

\bibitem{meng2023topological}
\bibinfo{author}{Meng, C.}, \bibinfo{author}{Wu, J.-S.} \& \bibinfo{author}{Smalyukh, I.~I.}
\newblock \bibinfo{title}{Topological steering of light by nematic vortices and analogy to cosmic strings}.
\newblock \emph{\bibinfo{journal}{Nat. Mater.}} \textbf{\bibinfo{volume}{22}}, \bibinfo{pages}{64--72} (\bibinfo{year}{2023}).

\bibitem{PhysRevA.88.041803}
\bibinfo{author}{Sheng, J.}, \bibinfo{author}{Miri, M.-A.}, \bibinfo{author}{Christodoulides, D.~N.} \& \bibinfo{author}{Xiao, M.}
\newblock \bibinfo{title}{$\mathcal{PT}$-symmetric optical potentials in a coherent atomic medium}.
\newblock \emph{\bibinfo{journal}{Phys. Rev. A}} \textbf{\bibinfo{volume}{88}}, \bibinfo{pages}{041803} (\bibinfo{year}{2013}).

\bibitem{feynman1957geometrical}
\bibinfo{author}{Feynman, R.~P.}, \bibinfo{author}{Vernon~Jr, F.~L.} \& \bibinfo{author}{Hellwarth, R.~W.}
\newblock \bibinfo{title}{Geometrical representation of the schr{\"o}dinger equation for solving maser problems}.
\newblock \emph{\bibinfo{journal}{J. Appl. Phys.}} \textbf{\bibinfo{volume}{28}}, \bibinfo{pages}{49--52} (\bibinfo{year}{1957}).

\end{thebibliography}

\begin{addendum}
\item We thank A. Agbo Bidi for useful discussions.
This work was supported by National Natural Science Foundation of China (No.52488201, and No.12074306) and European Union's Horizon 2020 program, through a FET Open research and innovation action under the grant agreement No. 964770 (TopoLight). Additional support was provided by the ANR Labex GaNext (ANR-11-LABX-0014), the ANR program "Investissements d'Avenir" through the IDEX-ISITE initiative 16-IDEX-0001 (CAP 20-25), the ANR project MoirePlusPlus, and the ANR project "NEWAVE" (ANR-21-CE24-0019). IS acknowledges support from the Deutsche Forschungsgemeinschaft (DFG, German Research Foundation, project numbers 447948357 and 440958198),
the Sino-German Center for Research Promotion (Project M-0294), the ERC (Consolidator Grant 683107/TempoQ), the German Ministry of Education and Research (Project QuKuK, BMBF Grant No. 16KIS1618K) and the Stiftung der
Deutschen Wirtschaft.
\item[Competing interests] The authors declare no competing interests.
\item[Correspondence] Correspondence
should be addressed to Z. Zhang (zhyzhang@xjtu.edu.cn), Ismael Septembre (ismael.septembre@uni-siegen.de), G. Malpuech (guillaume.malpuech@uca.fr), \\
D. Solnyshkov (dmitry.solnyshkov@uca.fr).

\end{addendum}

\renewcommand{\figurename}{Extended Data Figure}
\setcounter{figure}{0}
\clearpage
\section*{Extended Data Figures}

\begin{figure}[b]
\centering
\includegraphics[width=0.5\linewidth]{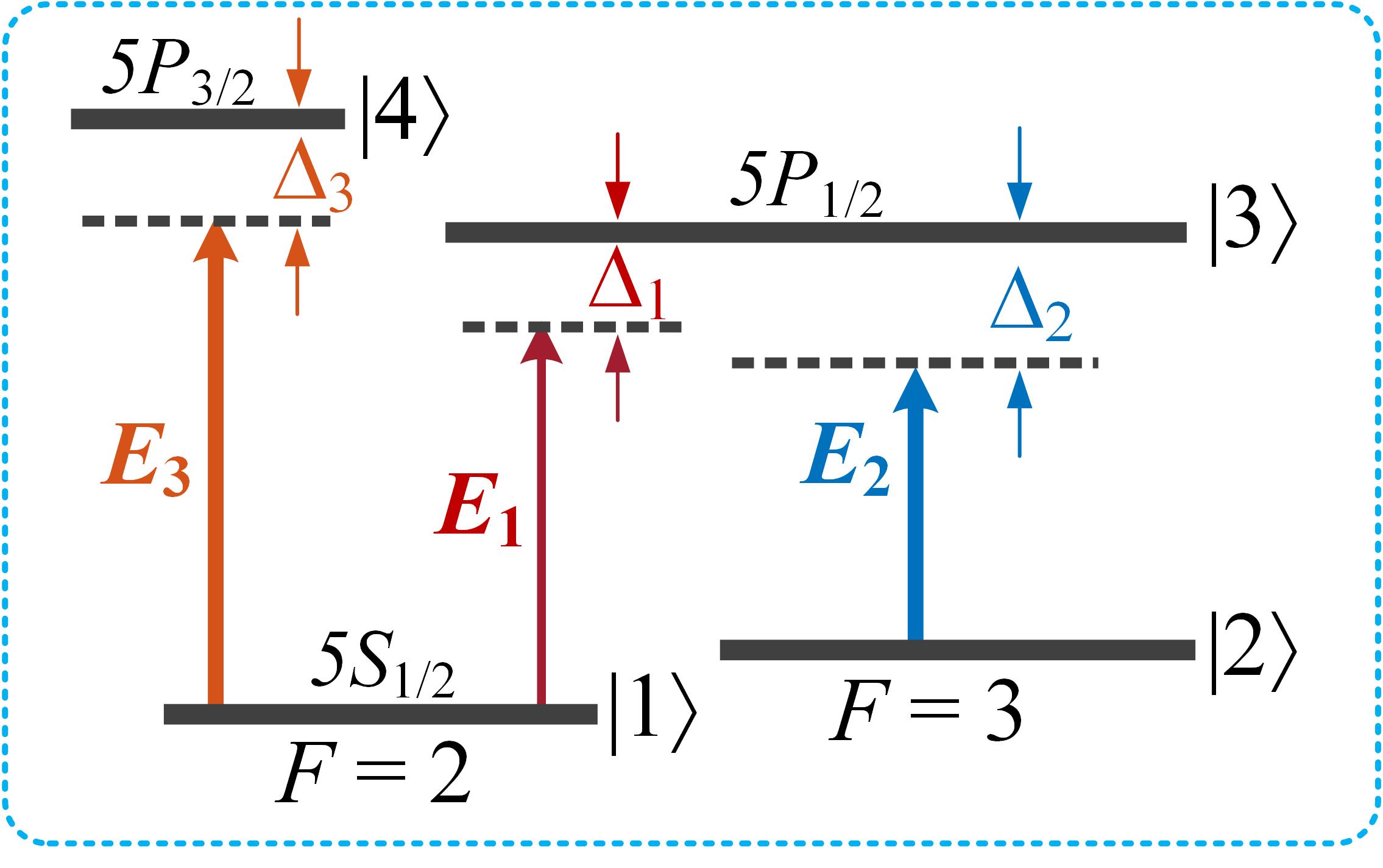}
\caption{\textbf{The four-level $N$-type $^{85}Rb$ atomic configuration.} \label{figS1}}
\end{figure}

\begin{figure}[tbp]
\centering
\includegraphics[width=0.9\linewidth]{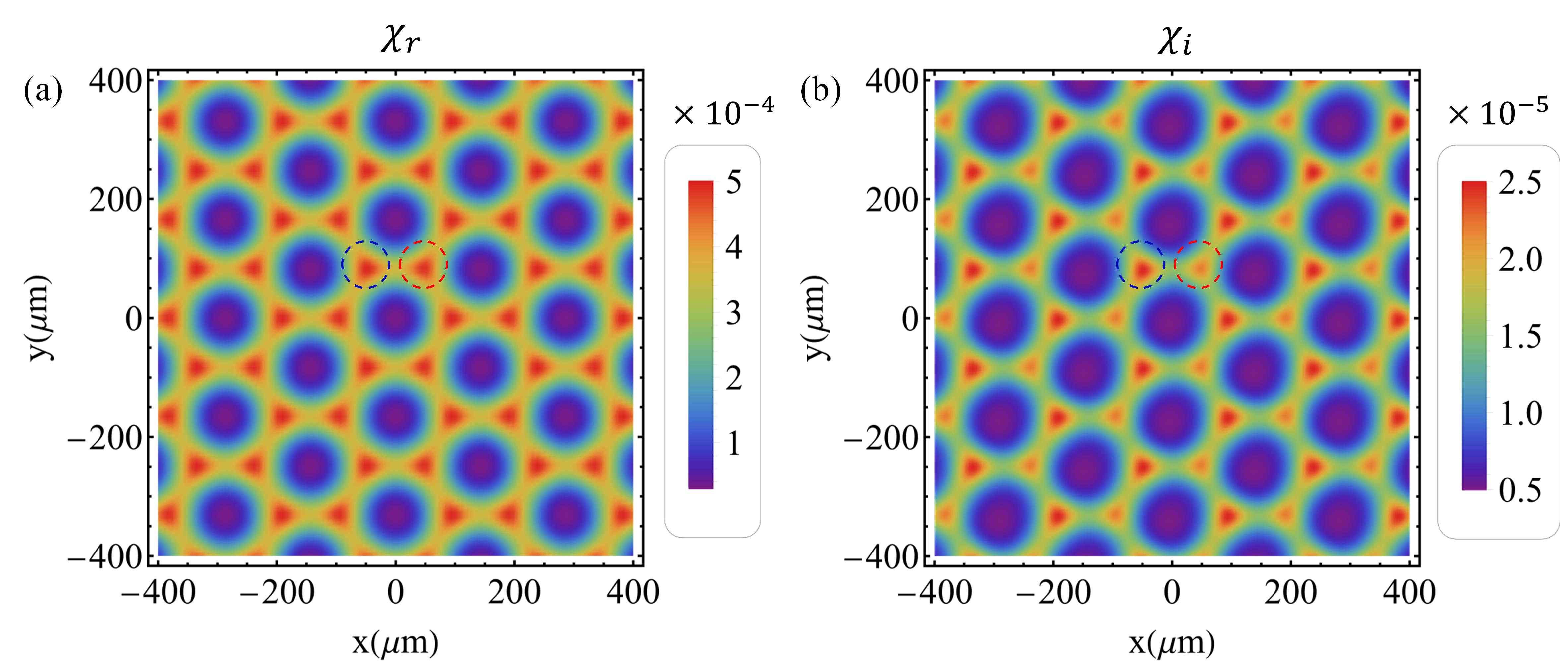}
\caption{\textbf{Simulated susceptibility.} Real \textbf{(a)} and imaginary \textbf{(b)} part distribution of susceptibility in the honeycomb photonic lattice constructed in a four-level atomic system.} \label{figS2}
\end{figure}

\begin{figure}[tbp]
\centering
\includegraphics[width=0.99\linewidth]{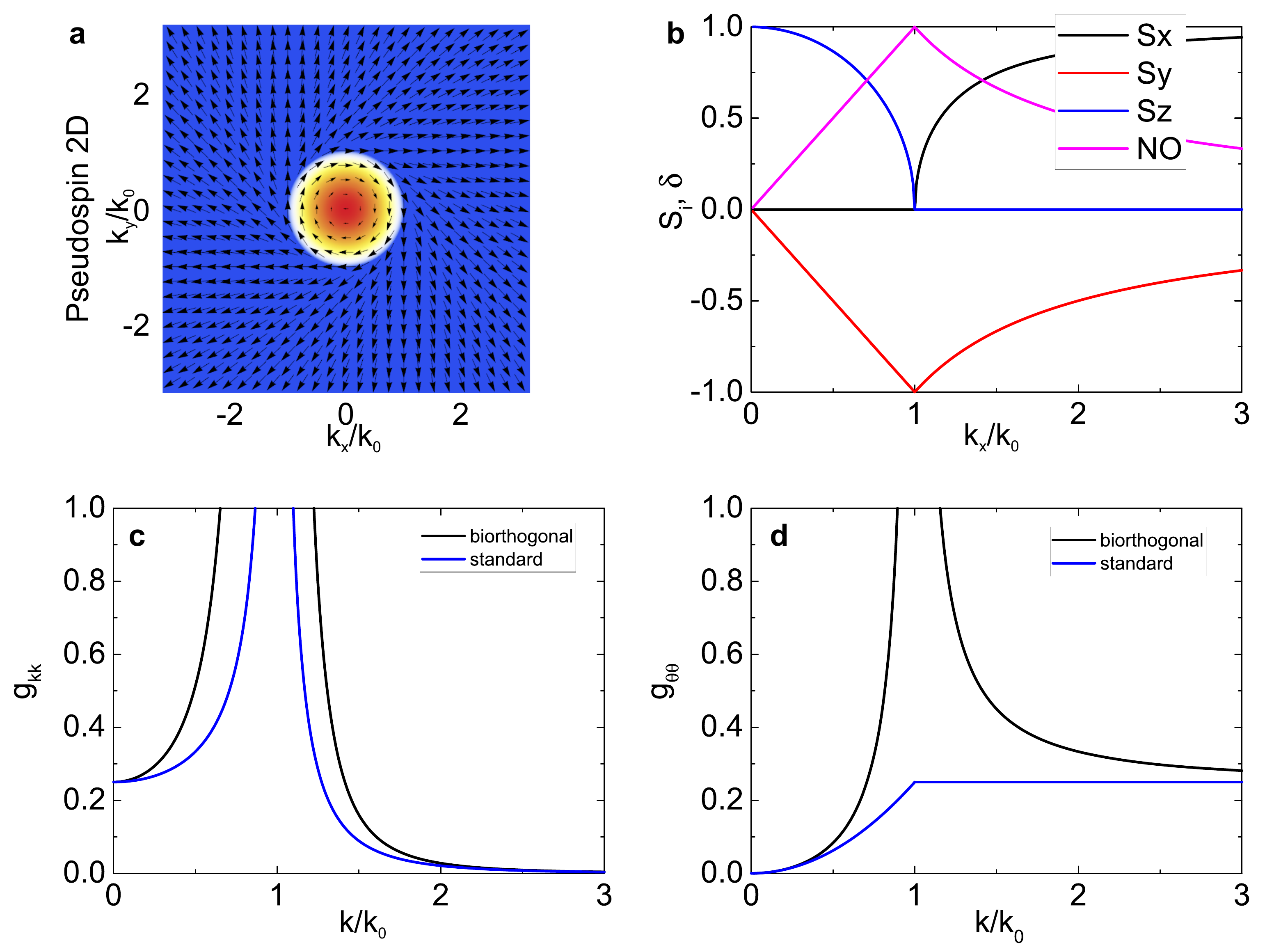}
\caption{\textbf{The eigenstates and the quantum metric.}
\textbf{a} The 2D distribution of the pseudospin of the eigenstates for the exceptional ring. \textbf{b} Radial distribution of the pseudospin components $(S_x,S_y,S_z)$ of the eigenstates. Magenta line - non-orthogonality of the two eigenstates $\delta= \left | \left \langle \psi_+\middle | \psi_-\right \rangle \right |$. \textbf{c} The radial quantum metric $g_{kk}$: biorthogonal (black) and standard (blue). \textbf{d} The azimuthal quantum metric $g_{\theta\theta}$: biorthogonal (black) and standard (blue).
\label{EDFpseudoNOmetric}}
\end{figure}

\begin{figure}[tbp]
\centering
\includegraphics[width=0.99\linewidth]{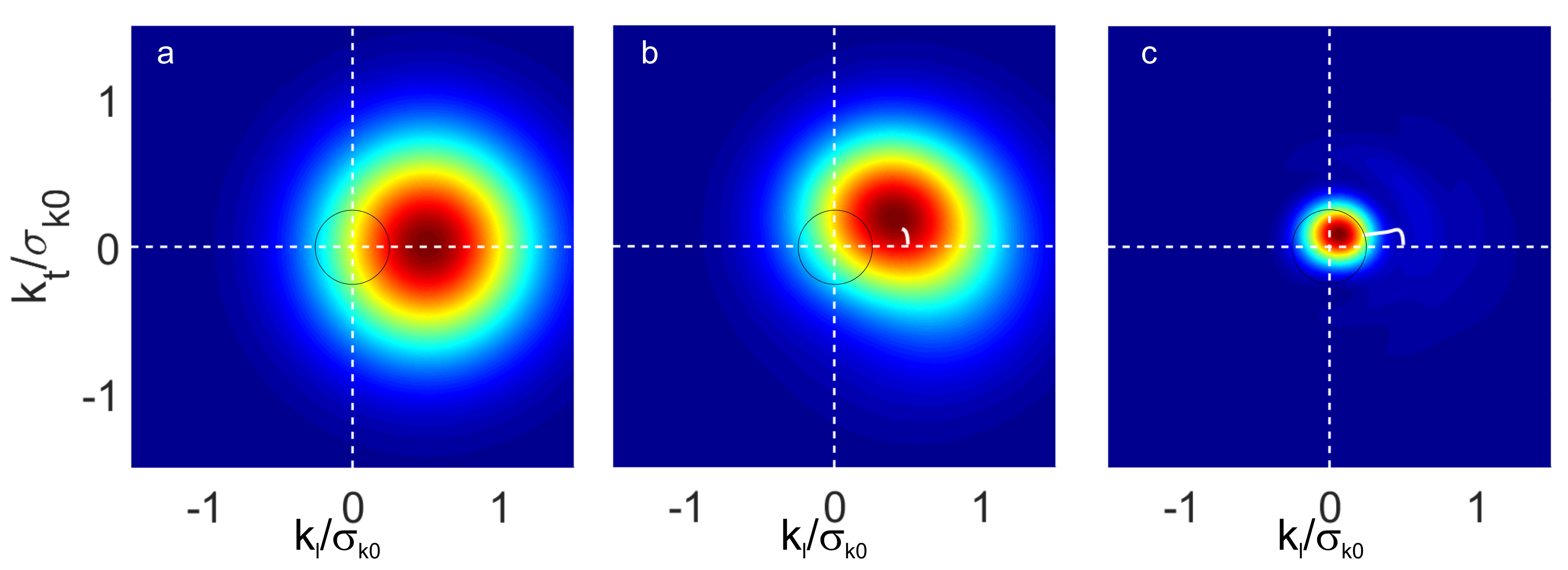}
\caption{\textbf{Time snapshots of numerically simulated non-Hermitian drift.} 
\textbf{a}-\textbf{c} The probability density $|\psi(x,y,t)|^2$ (equivalent to the intensity of the electric field) at different moments of time $t/t_0=0,0.25,1$ (equivalent to propagation length in the paraxial approximation), where $t_0$ corresponds to the length of the experimental vapor cell. White curve marks the trajectory of the center of mass over time. Black circle marks the exceptional ring. \label{EDF1snap}}
\end{figure}

\begin{figure}[tbp]
\centering
\includegraphics[width=0.99\linewidth]{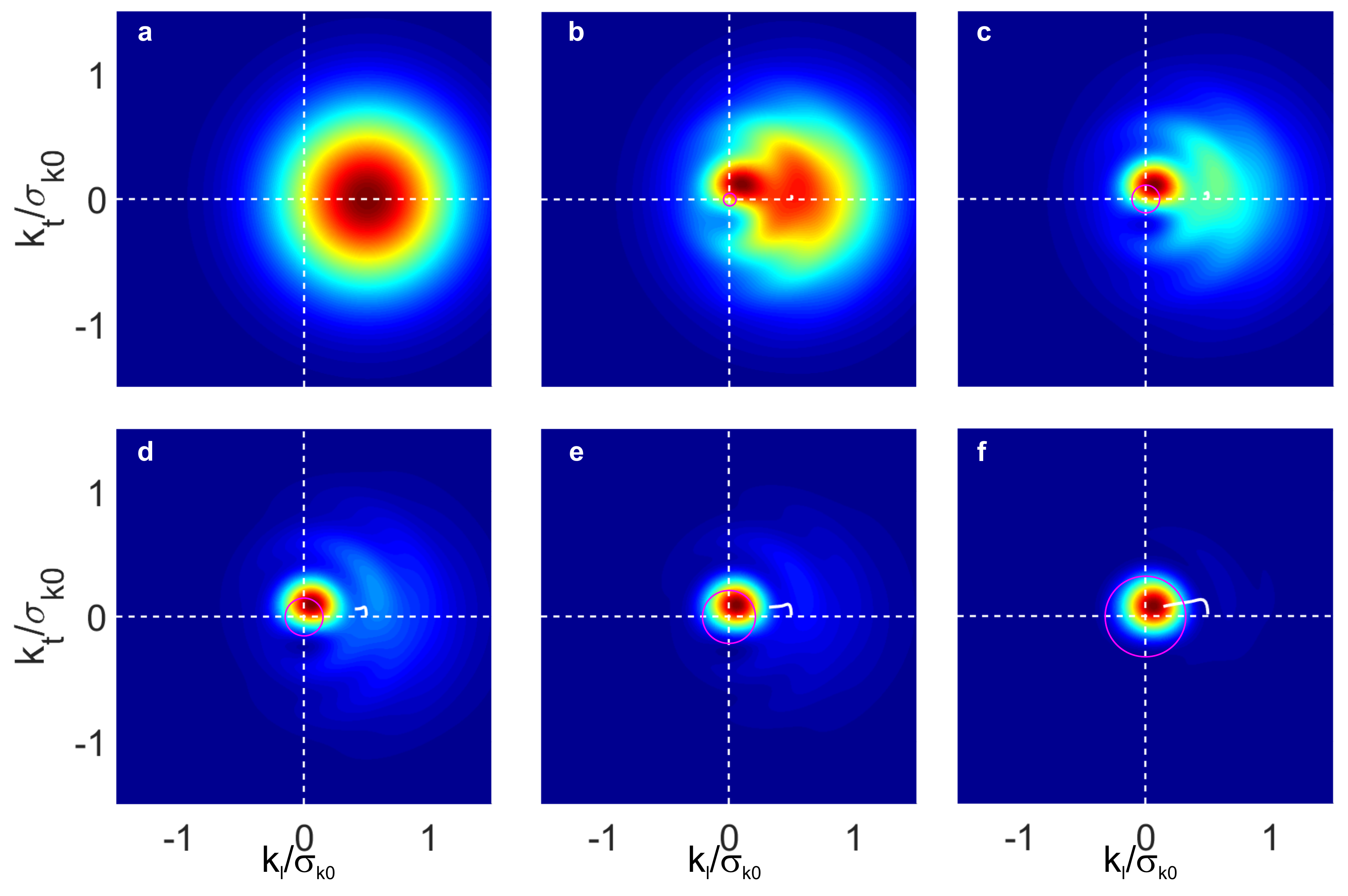}
\caption{\textbf{Numerically simulated non-Hermitian drift for varying non-Hermiticity.} 
\textbf{a}-\textbf{f} The probability density $|\psi(x,y,t_0)|^2$ (equivalent to the intensity of the electric field) at the final moment of time for different values of $\gamma$ and different exceptional ring sizes (magenta circles). White curve marks the trajectory of the center of mass over time in each case. \label{EDF2gam}}
\end{figure}




\end{document}